\documentclass[aps,prl,10pt,twocolumn,floatfix,showpacs,superscriptaddress,longbibliography]{revtex4-1}

\usepackage{graphicx,amsmath,amsfonts,bm}
\usepackage[colorlinks=true, citecolor=blue, linkcolor=blue, urlcolor=blue]{hyperref}
\usepackage[all]{hypcap}

\begin{document}

\title{$\mathbb{Z}_4$ parafermions in weakly interacting superconducting constrictions at the helical edge of quantum spin Hall insulators}
\author{C. Fleckenstein}
\email{christoph.fleckenstein@physik.uni-wuerzburg.de}
\affiliation{Institute of Theoretical Physics and Astrophysics, University of W\"urzburg, 97074 W\"urzburg, Germany}
\author{N. Traverso Ziani}
\affiliation{Institute of Theoretical Physics and Astrophysics, University of W\"urzburg, 97074 W\"urzburg, Germany}
\affiliation{Dipartimento di Fisica, Universit\`a di Genova, 16146 Genova, Italy}

\author{B. Trauzettel}
\affiliation{Institute of Theoretical Physics and Astrophysics, University of W\"urzburg, 97074 W\"urzburg, Germany}
\begin{abstract}
Parafermions are generalizations of Majorana fermions that may appear in interacting topological systems. They are known to be powerful building blocks of topological quantum computers. Existing proposals for realizations of parafermions typically rely on strong electronic correlations which are hard to achieve in the laboratory. We identify a novel physical system in which parafermions generically develop. It is based on a quantum point contact formed by the helical edge states of a quantum spin Hall insulator in vicinity to an ordinary $s$-wave superconductor. Interestingly, our analysis suggests that $\mathbb{Z}_4$ parafermions are emerging bound states in this setup -- even in the {\it weakly interacting} regime.
\end{abstract}

\maketitle

\textit{Introduction.--} During the last decades, topological quantum physics has become one of the most active directions of modern condensed matter research. Especially, the formation of topological boundary excitations, such as Majorana fermions \cite{EMajorana, Kitaev2001}, has attracted a lot of attention, both theoretically as well as experimentally \cite{vonOppen2010, Lutchyn2010,TPchoy2011,Mourik2012,Rokhinson2012,SNadj2013,SNadj2014,Marcus2016}. These robust bound states have been proposed in various host materials, ranging from vortices in $p_x+ip_y$ superconductors \cite{Volovik1999,NRead2000} over ferromagnet-superconductor heterojunctions in quantum spin Hall insulators (QSHIs) \cite{KaneMele2005, KaneMele2005b, Bernevig2006, Konig, Roth2009, LFu2008, Akhmerov2009} to spin-orbit coupled quantum wires \cite{vonOppen2010,Lutchyn2010}. Due to their non-Abelian statistics \cite{Nayak2008,Ivanov2001, Alicea2011}, the interest in those topological bound states is not only fundamental but also practical: They can potentially be used for protocols in topological quantum computation (TQC) \cite{Kitaev2003}. Majorana fermions are the conceptually simplest representatives of non-Abelian particles. However, braiding of Majorana fermions is not able to generate all the operations needed for universal TQC. For this task, more complex anyonic particles, assigned in general to a $\mathbb{Z}_n$ permutation group, are required \cite{Nayak2008,Hutter2016}. Due to the high groundstate degeneracy of those $\mathbb{Z}_n$ anyons, electron-electron interactions are essential in physical realizations thereof. In particular, $\mathbb{Z}_n$ parafermions are concrete examples of topological states that are proposed to emerge in correlated topological systems.

Recently, possible realizations of those exotic bound states have been predicted in different setups, including interacting QSHIs \cite{FZhang2014,Orth2015,CChen2016,YVinkler2017}, fractional quantum Hall insulators \cite{HLindner2012,DJClarke2013,AVaezi2013,AVaezi2014,
YAlavirad2017}, fractional QSHIs \cite{Klinovaja2014}, quantum wires \cite{Klinovaja2014b}, or lattice systems \cite{LMazza2018,AChew2018,ACalzona2018}. Typically, $s$-wave superconductors are placed in proximity to a repulsively interacting region of the electronic system. Then, parafermionic bound states can form at the interface between two distinct regions in space. The experimental realization of parafermions is however an unsolved and undoubtedly challenging task. Difficulties arise as superconductors and strong magnetic fields are, for instance, required at the same time in fractional quantum Hall systems. In QSHI, where magnetic fields are not essential, many proposals rely on particularly strong repulsive interactions at the corresponding helical edge. Although topological insulators based on InAs/GaSb quantum wells \cite{IKnez2011, Spanton2014, LJDu2015} have been shown to present a platform for repulsively interacting helical edge states, the magnitude of the interaction strength consistent with experimental data is under debate \cite{TLi2015, Geissler2016}.

From our point of view, a feasible proposal for the generation of parafermions in the laboratory is still lacking.
\begin{figure}
\centering
\includegraphics[scale=0.5]{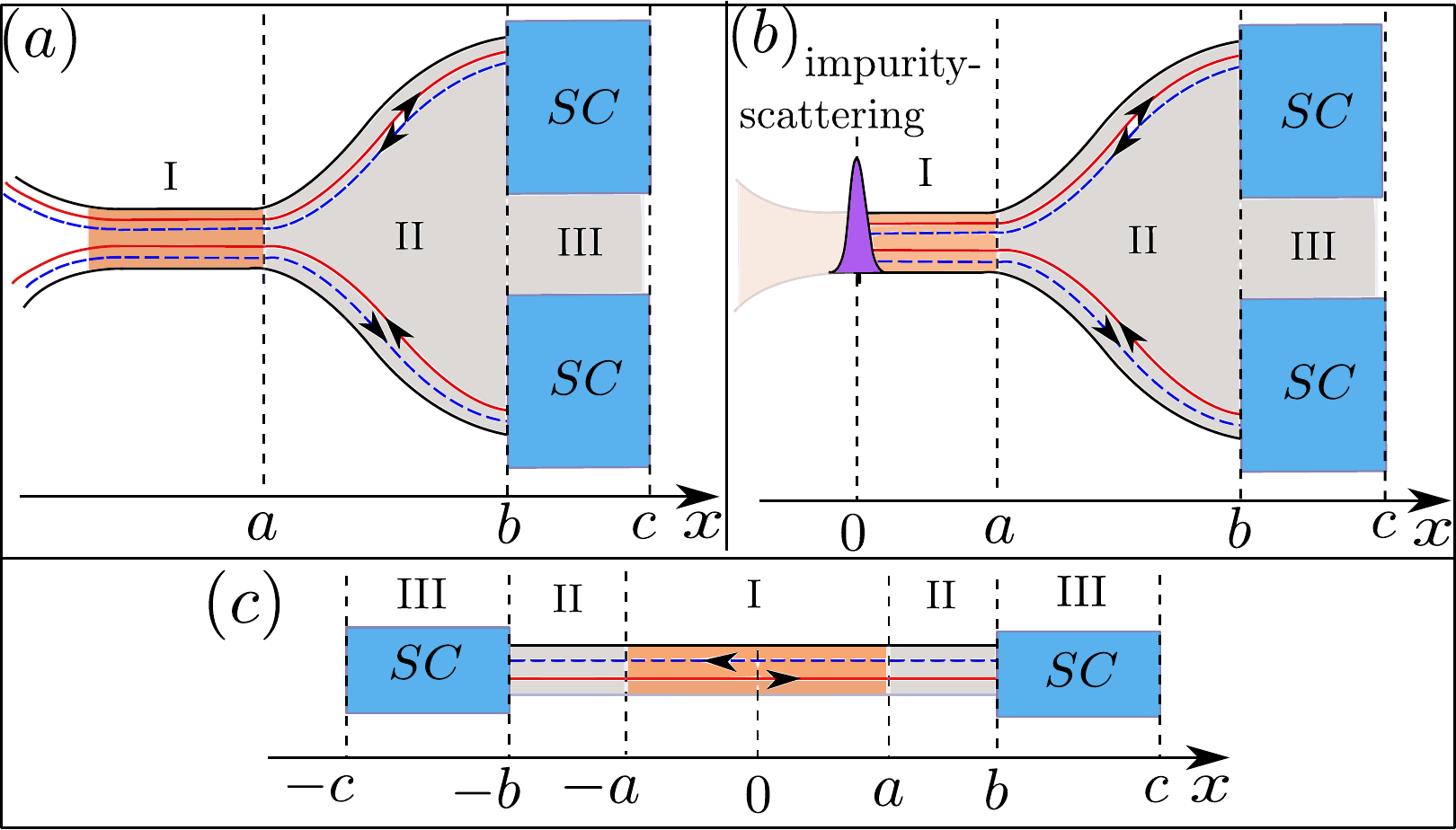}
\label{Fig:system}
\caption{(a) Schematic of the system: an extended constriction in a stripe of a 2D topological insulator is brought in vicinity to two regions with proximity induced superconductivity. (b) The same setup as in (a) with an additional impurity that totally pinches off the constriction. (c) Schematic of the unfolded structure after applying appropriate boundary conditions to case (b).}
\end{figure}
We argue to close this gap in this work. The system we propose is a quantum point contact at the (weakly) interacting helical edge of a QSHI in proximity to two ordinary $s$-wave superconductor (see Fig.~\ref{Fig:system}(a) for a schematic). We are inspired by the investigation of similar setups in the absence of electronic correlations. In particular, the formation of Majorana bound states \cite{Lutchyn2016} and the emergence of odd-frequency superconductivity \cite{Fleckenstein2018} has been theoretically proposed.

In the presence of interactions, the system becomes evidently much richer. Indeed, the constriction gives rise to several interaction terms that are relevant in the renormalization group (RG) sense for a wide range of repulsive interactions. Generically, single- and two-particle scattering terms have to be taken into account. For the appearance of parafermions, two-particle scattering has to dominate over single-particle scattering. Surprisingly, we argue below that this can even happen in the weakly interacting regime in our system. We identify two distinct cases schematically illustrated in Fig.~\ref{Fig:system}(a) and (b), respectively. In case (a), a Majorana bound state and a parafermion coexist in region II in a non-local fashion, i.e. across the two edges of the QSHI. In case (b), where the constriction is totally pinched off, which we illustrate by an impurity in the figure, two local parafermions appear in region II being spatially separated at the two edges of the QSHI.

\textit{Model.--} The starting point of our analysis are the two helical edge states formed at the boundary of a QSHI. With $\hbar=1$, the kinetic energy is then described by the fermionic Hamiltonian
\begin{eqnarray}
\label{Eq:H0}
H_0=\int\!\mathrm{d}x\!\sum_{\substack{l=(R,L)=(+,-) \\ \sigma=\uparrow,\downarrow}}\!\!\hat{\psi}_{l,\sigma}^{\dagger}(x)(-iv_Fl\partial_x)\hat{\psi}_{l,\sigma}(x)
\end{eqnarray}
with the Fermi field operators of the upper ($\hat{\psi}_{R,\uparrow}(x),~\hat{\psi}_{L,\downarrow}(x)$) and the lower edge ($\hat{\psi}_{R,\downarrow}(x),~\hat{\psi}_{L,\uparrow}(x)$), respectively. Including density-density interactions in the usual way, we can bosonize the theory exploiting the bosonization identity in the charge-spin basis \cite{TGiamarchi2003}, i.e.
\begin{equation}
\label{Eq:FermiField}
\hat{\psi}_{r,\nu}(x)=\frac{\hat{U}_{r,\nu}e^{irk_Fx}}{\sqrt{2\pi\alpha}}e^{-\frac{i}{\sqrt{2}}\left[r\phi_{\rho}(x)-\theta_{\rho}(x)+\nu\left(r\phi_{\sigma}(x)-\theta_{\sigma}(x)\right)\right]},
\end{equation}
where $r=R,L=+,-$ and $\nu=\uparrow,\downarrow=+,-$. $\hat{U}_{r,l}$ are Klein factors lowering the number of Fermions by one. In Eq.~(\ref{Eq:FermiField}), $\alpha$ denotes a high-energy cutoff. The conjugate bosonic fields $\phi_{\rho/\sigma}(x),~\theta_{\rho/\sigma}(x)$ are linear combinations of bosonic fields on the upper and lower edge (designated by the indices $1$ and $2$):
$\phi_{\rho}=1/\sqrt{2}(\phi_{1}(x)+\phi_2(x))$, $\phi_{\sigma}=1/\sqrt{2}(\theta_2(x)-\theta_1(x))$, $\theta_{\rho}=1/\sqrt{2}(\theta_{1}(x)+\theta_2(x))$, $\theta_{\sigma}=1/\sqrt{2}(\phi_{2}(x)-\phi_1(x))$, obeying the commutation relations
\begin{equation}
\label{Eq:commutator}
\left[\phi_{\nu}(x),\theta_{\mu}(y)\right]=i\pi~\mathrm{\theta}(y-x)\delta_{\nu\mu}.
\end{equation}
The interacting extension of the Hamiltonian of Eq. (\ref{Eq:H0}) can then be written in the well-known bosonized form
\begin{eqnarray}
H_0\!=\!\!\frac{1}{2\pi}\!\!\int\!\!\! \mathrm{d}x\!\!\!\sum_{\nu=\rho,\sigma}\!\!\left[\!\frac{u_{\nu}}{K_{\nu}}\!\left(\partial_x\phi_{\nu}(x)\right)^2\!+\!u_{\nu}\!K_{\nu}\!\left(\partial_x \theta_{\nu}(x)\right)^2\!\right]
\end{eqnarray}
with renormalized velocities $u_{\nu}$ and Luttinger interaction parameters $K_{\rho}$ and $K_{\sigma}$ characterizing the interaction strength. For helical Luttinger liquids, where spin-rotation invariance is strongly broken, $K_{\rho}<1$ and $K_{\sigma}> 1$ for repulsive interactions, likewise, $K_{\rho}>1$ and $K_{\sigma}<1$ for attractive interactions. For vanishing inter-edge interaction strength, the interaction parameters of both channels (charge and spin) are even coupled to each other by $K_{\rho}=1/K_{\sigma}\equiv K_0$ \cite{Hou2009}. This strong coupling of $K_{\rho}$ and $K_{\sigma}$ is, however, lost if inter-edge interaction is switched on. Hence, in our system, interaction parameters should obey a spatial dependence when the two helical edges of the QSHI are brought together in the constriction. There, we expect to have $K_{\rho}<K_0$ and $1\leq K_{\sigma}<1/K_0$ provided that intra-edge interactions are stronger than inter-edge interactions.

Apart from density-density interactions, in regions I and III of Fig.~\ref{Fig:system}(a), additional interaction terms, that do not result in a quadratic form after Bosonization, have to be taken into account. In region III, we consider superconducting $s$-wave pairing. This can be incorporated on the basis of a BCS mean field approach by the following fermionic Hamiltonion
\begin{equation}
\label{Eq:Hsc}
H_{\Delta}=\!\!\int\! \mathrm{d}x \Delta(x)\!\left[\hat{\psi}_{R,\!\uparrow}^{\dagger}(x)\hat{\psi}_{L,\downarrow}^{\dagger}(x)+\hat{\psi}_{L,\!\uparrow}^{\dagger}(x)\hat{\psi}_{R,\downarrow}^{\dagger}(x)\!\right]\!+\!\mathrm{h.c.}\, ,
\end{equation}
where $\Delta(x)$ is a spatially dependent pairing potential. Since we do not assume a connection between the two helical edges in region III, the corresponding Hamiltonian is diagonal in the fields of upper and lower edge. Using the bosonization identity (\ref{Eq:FermiField}) neglecting Klein factors \cite{TGiamarchi2003}, the bosonized form of Eq.~(\ref{Eq:Hsc}) becomes
\begin{equation}
H_{\Delta}=\int \mathrm{d}x \tilde{\Delta}(x)\lbrace \sin\left[2\theta_1(x)\right]+\sin\left[2\theta_2(x)\right]\rbrace
\end{equation}
with $\tilde{\Delta}(x)=\Delta(x)/(\pi\alpha)$.

For the constriction in region I, we consider all possible single- and two-particle scattering terms that (i) preserve time-reversal symmetry, (ii) are able to open a (partial) gap, and (iii) are relevant in the RG sense for a wide range of (weak) repulsive interactions, see Supplementary Material (SM). Those terms are in fermionic representation
\begin{eqnarray}
\label{Eq:Hsing}
H_{\mathrm{t}} &=& \int \mathrm{d}x t(x)\left[\hat{\psi}^{\dagger}_{R,\!\uparrow}\hat{\psi}_{L,\uparrow}+\hat{\psi}^{\dagger}_{L,\downarrow}\hat{\psi}_{R,\downarrow}\right]+\mathrm{h.c.},\\
\label{Eq:Hum}
H_{\mathrm{um}} &=& \int \mathrm{d}x g_{\mathrm{um}}(x)\hat{\psi}_{R,\uparrow}^{\dagger}\hat{\psi}_{R,\downarrow}^{\dagger}\hat{\psi}_{L,\downarrow}\hat{\psi}_{L,\uparrow}+\mathrm{h.c.},\\
\label{Eq:Hpbs}
H_{\mathrm{pbs}} &=& \int \mathrm{d}xg_{\mathrm{pbs}}(x)\hat{\psi}_{R,\uparrow}^{\dagger}\hat{\psi}_{L,\downarrow}\hat{\psi}_{L,\uparrow}^{\dagger}\hat{\psi}_{R,\downarrow}+\mathrm{h.c.},
\end{eqnarray}
where we have dropped the explicit spatial dependence of the field operators to save space.
Note that the single-particle scattering term (\ref{Eq:Hsing}) \cite{Lutchyn2016, JCYTeo2009, CXLiu2011, FDolcini2011, GDolcetto2012} and the Umklapp scattering term (\ref{Eq:Hum}) \cite{TGiamarchi2003} are spin-preserving processes, while the pair backscattering term (\ref{Eq:Hpbs}) \cite{VGritsev2005, MCheng2011,Seldmayr2013} requires breaking of axial spin symmetry \cite{CWu2006}. Applying the bosonization identity (\ref{Eq:FermiField}) to
Eqs.~(\ref{Eq:Hsing}-\ref{Eq:Hpbs}) neglecting Klein factors, we obtain the bosonized Hamiltonians
\begin{eqnarray}
\label{Eq:HsingA}
&& \!\!\!\! H_{\mathrm{t}}= \int \mathrm{d}x\tilde{t}(x) \!\cos[\sqrt{2}\phi_{\rho}(x)-2k_Fx]\cos[\sqrt{2}\phi_{\sigma}(x)],\\
\label{Eq:HumA}
&& \!\!\!\! H_{\mathrm{um}}= \int \mathrm{d}x\tilde{g}_{\mathrm{um}}(x)\cos[2\sqrt{2}\phi_{\rho}(x)-4k_Fx],\\
\label{Eq:HpbsA}
&& \!\!\!\! H_{\mathrm{pbs}}= \int \mathrm{d}x  \tilde{g}_{\mathrm{pbs}}(x)\cos[2\sqrt{2}\theta_{\sigma}(x)]
\end{eqnarray}
with $\tilde{t}(x)=2t(x)/(\pi\alpha)$, $\tilde{g}_{\mathrm{s}}(x)=g_{\mathrm{s}}(x)/(2\pi^2\alpha^2)$ and $\tilde{g}_{\mathrm{pbs}}(x)=g_{\mathrm{pbs}}(x)/(2\pi^2\alpha^2)$. It is important to notice for our subsequent analysis that $H_{\mathrm{um}}$ and  $H_{\mathrm{pbs}}$ commute with each other but both do not commute with $H_{\mathrm{t}}$ \cite{sup1}. Hence, they cannot be ordered simultaneously in the same region of space. Therefore, pinning of the bosonic fields in the strong coupling regime is only possible if either single- or two particle scattering dominates the physics.
For the emergence of parafermions, it is mandatory that at least one of the two two-particle scattering terms ($H_{\mathrm{um}}$ or $H_{\mathrm{pbs}}$) provides the dominant interaction. Only then the required groundstate degeneracy is present.

\begin{figure}
\centering
\includegraphics[scale=0.4]{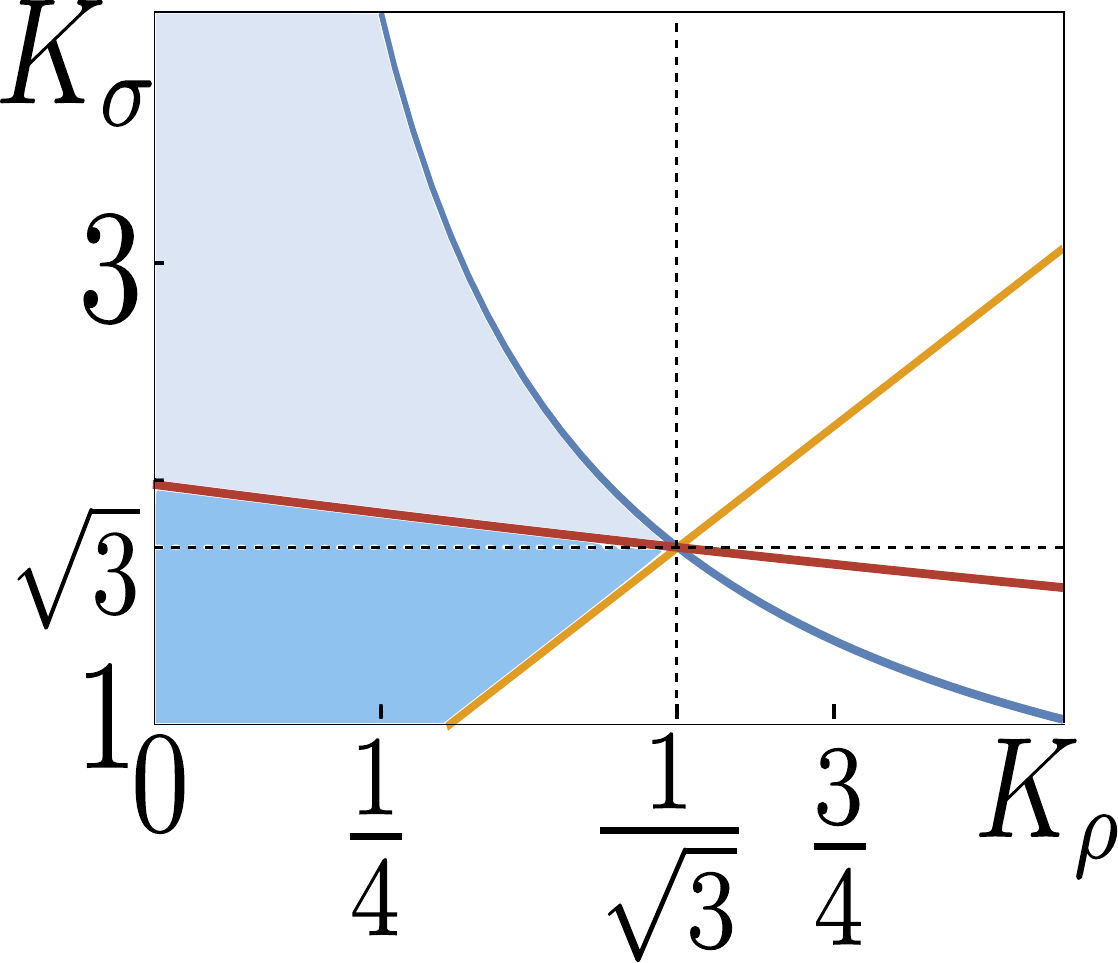}
\label{Fig:RG}
\caption{Illustration of the conditions obtained from the RG analysis of the various terms. The lines correspond to $K_{\sigma}=3K_{\rho}$ (orange), $K_{\sigma}=-K_{\rho}/2+\sqrt{K_{\rho}^2/4+16}$ (red), $K_{\sigma}=1/K_{\rho}$ (blue). The light and dark blue shaded areas mark the parameter regime where two-particle terms are RG-dominant.}
\end{figure}

We consider two measures for the subsequent discussion of the relative importance of the terms in Eqs.~(\ref{Eq:HsingA}-\ref{Eq:HpbsA}): (i) RG arguments and (ii) averaging due to oscillations of the integrand at finite $k_F$. At the Dirac point, where $k_F=0$, the measure (ii) does not apply and only RG arguments count. The corresponding RG equations of the terms (\ref{Eq:HsingA}-\ref{Eq:HumA}) are discussed in the SM. It is shown that they are all three RG-relevant for a wide range of repulsive interactions \cite{MCheng2011, Dolcetto2016, KaneFisher1992, Schmidt2017}. We can order their relative importance according to their scaling dimension. This ordering yields the following inequalities between $K_{\rho}$ and $K_{\sigma}$ for the regime in parameter space where two-particle terms should dominate the low-energy physics
\begin{eqnarray}
\label{Eq:RG1a}
K_{\sigma}&>&3K_{\rho},\\
\label{Eq:RG2a}
K_{\sigma}&>&-K_{\rho}/2+\sqrt{K_{\rho}^2/4+4}.
\end{eqnarray}
When Eq.~(\ref{Eq:RG1a}) (Eq.~(\ref{Eq:RG2a})) is fulfilled, $H_{\mathrm{um}}$ ($H_{\mathrm{pbs}}$) dominates over $H_{\mathrm{t}}$. The conditions are illustrated in Fig.~\ref{Fig:RG}. As $K_{\sigma}$ cannot exceed $1/K_{\rho}$ in our model, the shaded areas represent the parameter space for which at least one of the two two-particle processes is dominant. For the emergence of parafermions, it is indeed sufficient (as shown below) that either $H_{\mathrm{pbs}}$ or $H_{\mathrm{um}}$ is more relevant than $H_{\mathrm{t}}$. In fact, when $H_{\mathrm{um}}$ is the most relevant interaction, we first pin $\phi_{\rho}$ to minimize the contribution from the dominating term $H_{\mathrm{um}}$. It turns out that for all pinned values that $\phi_{\rho}$ can take, $H_{\mathrm{t}}$ vanishes. Due to this property, we are subsequently also allowed to pin the bosonic field operators that characterize $H_{\mathrm{pbs}}$. Having $H_{\mathrm{pbs}}$ stronger than $H_{\mathrm{t}}$, and consequently pinning $\theta_\sigma$, this also allows us to neglect $H_{\mathrm{t}}$ on the basis of energetics \cite{Klinovaja2014}. The lowest energy states, which are degenerate in the absence of $H_{\mathrm{t}}$, are hence obtained by pinning $H_{\mathrm{um}}$. Taking into account that $K_{\rho}<K_0$, there is a parameter space, for which Eq.~(\ref{Eq:RG1a}) is satisfied, up to $K_0\sim 0.65$, see SM. This parameter regime still corresponds to rather strong interactions which does not seem to be a substantial gain compared to previous proposals for the emergence of parafermions.

However, at finite $k_F$, we discover a major advantage of our setup: $H_{\mathrm{t}}$ and $H_{\mathrm{um}}$ describe virtual transitions that cost energies $\sim 2E_F$ and $\sim 4E_F$, respectively, where $E_F=\hbar v_F k_F$ is the Fermi energy. If $k_F$ takes values of magnitude $(\pi/a)$ (or larger) with $a$ the length of the constriction, the importance of the terms $H_{\mathrm{t}}$ and $H_{\mathrm{um}}$ is expected to reduce, but not necessarily negligible. Therefore, the system remains to be gapped for some range of $k_F$. To conclude, for finite $k_F$, $H_{\mathrm{pbs}}$ can be the dominant term even for $1/\sqrt{3}<K_0<1$. The reason is that  $H_{\mathrm{pbs}}$ is the only term of Eqs.~(\ref{Eq:HsingA}-\ref{Eq:HpbsA}) in which the integrand does not oscillate at finite $k_F$. Let us emphasize that this argument holds in the weakly interacting regime which constitutes one of the main results of our work.

Under the assumption that two-particle scattering dominates the physics in the constriction at low energies for the reasons mentioned above, we drop the single-particle scattering completely for the subsequent discussion of parafermions. As the relative weight of $\tilde{g}_{\mathrm{um}}$ and $\tilde{g}_{\mathrm{pbs}}$ does neither influences the number, nor the position of the various minima (when pinning the bosonic fields), we choose for simplicity $g_{\mathrm{pbs}}(x)=g_{\mathrm{um}}(x)$ and reorganize Eqs. (\ref{Eq:HumA}) and (\ref{Eq:HpbsA}) into
\begin{equation}
\label{Eq:H2p}
H_{\mathrm{2p}}\!=\!H_{\mathrm{um}}+H_{\mathrm{pbs}}\!=\!\!\int \mathrm{d}x 2\tilde{g}_{\mathrm{pbs}}(x)\cos[2\phi_{1}(x)]\cos[2\phi_{2}(x)].
\end{equation}
This Hamiltonian constitutes the basis for the following construction of parafermionic operators.

\textit{Parafermions.--} Non-Abelian exchange statistics requires ground-state degeneracy. In the thermodynamic limit for the gapped phase of a sine-Gordon theory, it makes sense to assume that, deep inside the gapped area, the fields are pinned such that the corresponding cosine potential is minimized. The corresponding fields $\phi_{1/2}(x),~\theta_{1/2}(x)$ can take values $[0,2\pi[$ (modulo $2\pi$) \cite{FZhang2014}. Within this range, several minima of the assigned cosine-/sine-potentials can be reached, which implies a degenerate ground state. For the superconducting section (region III of Fig.~\ref{Fig:system}(a)), the fields $\theta_1(x)$ and $\theta_2(x)$ are pinned independently. To properly formalize this, we introduce phase fields $\hat{\theta}_{1}$ and $\hat{\theta}_{2}$, where the corresponding eigenvalues are designated to the pinned values. The gap induced by the constriction (region I of Fig.~\ref{Fig:system}(a)), however, involves both edges, implying a correlation between $\phi_1(x)$ and $\phi_2(x)$. Indeed, minimization of Eq. (\ref{Eq:H2p}) is achieved whenever one of the two cosines is maximized and the other is minimized. This constraint forces a relation between the assigned phase fields $\hat{\phi}_{1}$ and $\hat{\phi}_{2}$
\begin{equation}
\label{Eq:relation1}
\hat{\phi}_{2}=-\hat{\phi}_{1}-\pi/2+\pi \hat{l},
\end{equation}
where $\hat{\phi}_{1}$ takes the eigenvalues $\phi_{1}\in\lbrace 0,\pi/2,\pi,3/2\pi\rbrace$ and the integer valued operator $\hat{l}$ with eigenvalues $l\in\lbrace 1,2\rbrace$ (modulo 2) relates $\hat{\phi}_1$ and $\hat{\phi}_2$. The representation of $\phi_2$ in terms of $\phi_1$, together with Eq.~(\ref{Eq:commutator}) for the system shown in Fig.~\ref{Fig:system}(a), implies the following commutation relations
\begin{equation}
[\hat{l},\hat{\theta}_1]=[\hat{l},\hat{\theta}_2]=i,~~~
[\hat{\phi}_{1},\hat{\theta}_1]=i\pi,~~~[\hat{\phi}_1,\hat{\theta}_2]=0.
\end{equation}
With Eq.~(\ref{Eq:relation1}), the two-particle scattering (\ref{Eq:H2p}) in the constriction can be written as
\begin{equation}
H_{2p}=-\tilde{g}_{\mathrm{pbs}} a \left(\cos[4\hat{\phi}_{1}-2\pi\hat{l}]+\cos[2\pi\hat{l}]\right),
\end{equation}
where we assumed the length of each section to be $a$ for simplicity.

With the (quasi-)conjugate variables $\hat{\phi}_{1}$, $\hat{l}$, $\hat{\theta}_{1}$, $\hat{\theta}_{2}$ and the different sections being disjoint, we are able to construct parafermionic bound states at the interface between two neighboring sections, in a similar way as in Refs.~\cite{HLindner2012, DJClarke2013, Orth2015}. The major difference in our case (as compared to previous work) stems from the presence of the operator $\hat{l}$. It turns out that this operator can change the ground-state manifold and lead to {\it non-local} bound state operators. This non-locality arises as any parafermionic operator, applied to a certain ground-state, can not add energy to the system but rather projects the system onto another (degenerate) ground-state. This implies that the parafermionic creation operator necessarily needs to commute with the Hamiltonian. For our present case, however, it is not possible to write a purely local operator that obeys this constraint. This is indeed a direct consequence of the presence of the operator $\hat{l}$ that couples $\hat{\phi}_{1}$ and $\hat{\phi}_{2}$.

We find that the set of operators $\lbrace e^{i(\pi/2)\hat{S}},e^{i\pi\hat{l}}\rbrace$, where $\pi\hat{S}=\hat{\theta}_1-\hat{\theta}_2$, commute with the Hamiltonian and among themselfs. They induce the degenerate set of states $\vert s,l\rangle$, each of which satisfying $e^{i(\pi/2)\hat{S}}\vert s,l\rangle=e^{i(\pi/2)s}\vert s,l\rangle$ and $e^{i\pi\hat{l}}\vert s,l\rangle=e^{i\pi l}\vert s,l\rangle$ with distinct eigenvalues $s\in\lbrace 0,1,2,3\rbrace$ (modulo 4) and $l\in\lbrace 1,2\rbrace$ (modulo 2).
Moreover, it is easy to demonstrate that the operators
\begin{eqnarray}
\label{Eq:NonLocalParafermion}
\hat{\chi}_{s}=e^{i\pi/4}\!e^{i\pi\hat{\phi}_1}\!e^{i/2(\hat{\theta}_1-\hat{\theta}_2)}, ~~\hat{\chi}_l=e^{i\pi/2}\!e^{i/2(\hat{\theta}_1+\hat{\theta}_2)}\!e^{i\pi\hat{l}}~~~
\end{eqnarray}
commute with the Hamiltonian and describe creation operators of the quantum numbers $s$ and $l$, respectively,
\begin{eqnarray}
\label{Eq:NonLocalParafermion2}
\hat{\chi}_s\vert s,l\rangle\!=\!e^{i(\pi/2)s}\vert s+1,l\rangle, \hat{\chi}_{l}\vert s,l\rangle\!=\!e^{i\pi l}\vert s,l+1\rangle.
\end{eqnarray}
From Eqs.~(\ref{Eq:NonLocalParafermion}) and (\ref{Eq:NonLocalParafermion2}), we obtain the relations
\begin{equation}
\label{Eq:NonlocalParafermion3}
\hat{\chi}_s\hat{\chi}_l=e^{-i\pi/2}\hat{\chi}_l\hat{\chi}_s,~~~~\hat{\chi}_s^4=1,~~~~\hat{\chi}_l^2=1.
\end{equation}
These relations imply the simultaneous presence of a $\mathbb{Z}_4$ parafermion ($\hat{\chi}_s$) as well as a Majorana zero-mode ($\hat{\chi}_l$). To the best of our knowledge, this combination has not been predicted before. 
It should be noted that the fields $\hat{\theta}_1$ and $\hat{\theta}_2$ are necessarily contained within each operator. Hence, they describe non-local bound states delocalized across the upper and lower edge of region II in Fig.~\ref{Fig:system}(a).

Interestingly, we can slightly modify our setup according to Fig.~\ref{Fig:system}(b) to obtain local parafermions. The physical situation corresponds to a pinched off quantum point contact, which is either realized by a strong impurity or a physical boundary of the structure. We can model it by adding the impurity Hamiltonian \cite{KaneFisher1992}
\begin{equation}
\label{Eq:Imp}
H_{\mathrm{imp}} = V \left[\hat{\psi}^{\dagger}_{R,\uparrow}(0)\hat{\psi}_{L,\uparrow}(0) + \hat{\psi}^{\dagger}_{L,\downarrow}(0)\hat{\psi}_{R,\downarrow}(0) \right] + \mathrm{h.c.}.
\end{equation}
The corresponding (hard-wall) boundary conditions are derived on the fermionic level \cite{Fabrizio1995, Timm2012, Dolcetto2013, Fleckenstein2016a} (see SM). In bosonic language, it forces the following (non-local) relations between bosonic field operators
\begin{equation}
\label{Eq:BC4}
\hat{\phi}_2(x) = -\hat{\phi}_1(-x)-\pi/2, \;~~ \hat{\theta}_2(x) = \hat{\theta}_1(-x) .
\end{equation}
Although Eq.~(\ref{Eq:BC4}) contains two different points $x$ and $-x$ in space, the similarity to Eq.~(\ref{Eq:relation1}) is apparent, where now the operator $\hat{l}$ (that induced the non-locality of the resulting bound state operators) is absent.
Starting from the physical setup depicted in Fig.~\ref{Fig:system}(b), taking into account Eq.~(\ref{Eq:BC4}), we unfold the system to arrive at an equivalent (sine-Gordon) model illustrated in Fig.~\ref{Fig:system}(c). This sine-Gordon model contains the following mass terms
\begin{eqnarray}
\label{Eq:final1b}
H_{\Delta}\!&=&H_{\Delta-}\!+\!H_{\Delta+}\!=\!\tilde{\Delta}\!\bigg[\!\int_{\!-c\!}^{\!-b\!}\!\!+\!\int_{b}^{c}\bigg]\mathrm{d} x\sin  \![2\hat{\theta}_1(x)],\\
\label{Eq:final1c}
\tilde{H}_{\mathrm{2p}}\!&=&\!-\tilde{g}_{\mathrm{pbs}}\!\int_{-a}^a\!\!\mathrm{d}x \cos[2  \hat{\phi}_1(x)]\cos[2\hat{\phi}_1(-x)].
\end{eqnarray}
Minimization of Eq.~(\ref{Eq:final1c}) requires a constant field $\hat{\phi}_1(x)=\hat{\phi}_1(-x)=\hat{\phi}_1$ since any modulation with space adds energy $\propto \tilde{g}_{\mathrm{pbs}}$. With the introduction of phase fields $\hat{\phi}_{1}$ and $\hat{\theta}_{\pm}$ (where $\pm$ refers to the superconductor right ($+$) and left ($-$) of the origin in Fig.~\ref{Fig:system}(c)) we obtain effective Hamiltonians
\begin{eqnarray}
H_{\Delta\pm}&=&\tilde{\Delta}(c-b)\sin[2\hat{\theta}_{\pm}] ,\\
H_{\mathrm{2p},j}&=&-\tilde{g}_{\mathrm{pbs}} a (\cos[4\hat{\phi}_{1}]+1) .
\end{eqnarray}
The relevant bosonic field operators obey the following commutation relations $[\hat{\phi}_1,\hat{\theta}_-]=0$, $[\hat{\phi}_{1},\hat{\theta}_+]=i\pi$, which implies that
\begin{equation}
\hat{\xi}_-=e^{\hat{\phi}_1}e^{(i/2)\hat{\theta}_-},~~~\hat{\xi}_+=e^{i\hat{\phi}_1}e^{(i/2)\hat{\theta}_+}
\end{equation}
commute with the Hamiltonian and obey parafermionic exchange relations
\begin{equation}
\hat{\xi}_-\hat{\xi}_+=e^{-i\pi/2}\hat{\xi}_+\hat{\xi}_-,~~~\hat{\xi}_{\pm}^4=1.
\end{equation}
The four-fold degenerate ground-state manifold is formed by eigenstates of the operator $\hat{\xi}_-^{\dagger}\hat{\xi}_+=e^{(i/2)(\hat{\theta}_+-\hat{\theta}_-)}$, measuring the spin trapped in between the two superconducting regions in space. $\hat{\xi}_{\pm}$ are purely local operators, bound in region II of Fig.~\ref{Fig:system}(b), where each helical edge is occupied by a single parafermion. Including small overlaps between the two parafermionic bound states, this would result in a $8\pi$-periodic Josephson current, when a superconducting phase-shift is applied between the superconductors of upper and lower edge \cite{FZhang2014}.

To summarize, we have proposed a system composed of a quantum point contact and proximity induced s-wave superconductivity in a quantum spin Hall insulator that can host $\mathbb{Z}_4$ parafermionic bound states even in the weakly interacting regime. This finding is based on the competition between different coupling terms in the constriction. We discuss their relative importance and construct explicit operators for the bound states. Our predictions should be observable by tunneling spectroscopy or the Josephson effect.

\begin{acknowledgments} We acknowledge financial support by the DFG (SPP1666 and SFB1170 "ToCoTronics"), the ENB Graduate school on "Topological Insulators", and the Studienstiftung des Deutschen Volkes. We thank Alessio Calzona and Thomas Schmidt for interesting discussions.

\end{acknowledgments}

\section{Supplementary material}

\subsection{RG analysis in the constriction}
In the constricted area (region I of Fig.~\ref{Fig:system}(a) of the main text), Eqs.~(\ref{Eq:Hsing}-\ref{Eq:Hpbs})
can be shown to be relevant in the RG sense for the repulsively interacting regime.  Assuming that the couplings $\tilde{t}$, $\tilde{g}_{\mathrm{um}}$ and $\tilde{g}_{\mathrm{pbs}}$ are initially small, Eqs. (\ref{Eq:Hsing}-\ref{Eq:Hpbs}) can be analyzed in a perturbative RG approach \cite{MCheng2011, Dolcetto2016, KaneFisher1992}. In one-loop RG, we obtain
\begin{eqnarray}
\label{Eq:supRG1}
\frac{dy_{\mathrm{t}}}{dl}&=&(2-\frac{1}{2}(K_{\rho}+K_{\sigma}))y_{\mathrm{t}},\\
\label{Eq:supRG2}
\frac{dy_{\mathrm{um}}}{dl}&=&(2-2K_{\rho})y_{\mathrm{um}},\\
\label{Eq:supRG3}
\frac{dy_{\mathrm{pbs}}}{dl}&=&(2-2/K_{\sigma})y_{\mathrm{pbs}}
\end{eqnarray}
with the dimensionless coupling constants $y_{\mathrm{t}}=\tilde{g}_{\mathrm{t}}/(\pi u_{\sigma})$, $y_{\mathrm{um}}=\tilde{g}_{\mathrm{um}}/(\pi u_{\rho})$, $y_{\mathrm{pbs}}=\tilde{g}_{\mathrm{pbs}}/(\pi u_{\sigma})$, and the flow parameter $l$.
Eqs.~(\ref{Eq:supRG2}) and (\ref{Eq:supRG3}) imply that $y_{\mathrm{um}}$ and $y_{\mathrm{pbs}}$ are relevant in the whole repulsively interacting regime ($K_{\rho}<1$) provided $K_{\sigma}>1$. This is the case for our helical system as intra-edge interactions are typically stronger than inter-edge interactions. Given that $K_{\sigma}\leq 1/K_{\rho}$ and $K_{\rho}<1$, Eq.~(\ref{Eq:supRG1}) implies that $y_{\mathrm{t}}$ flows to strong coupling for $K_{\rho}>2-\sqrt{3}$. When $K_{\rho}<2-\sqrt{3}$, $y_{\mathrm{t}}$ is relevant provided $K_{\sigma}<4-K_{\rho}$.
In the regime, where all three terms are relevant in the RG sense, we order their relative importance according to their scaling dimension. As Umklapp and pair backscattering can be pinned simultaneously because they commute, it is particularly interesting to compare the scaling dimension of those two terms with the one of the single-particle scattering. On the basis of this comparison, we find that $H_{\mathrm{um}}$ and $H_{\mathrm{pbs}}$, respectively, dominate over $H_{\mathrm{t}}$ provided that
\begin{eqnarray}
\label{Eq:supRG1a}
K_{\sigma}&>&3K_{\rho},\\
\label{Eq:supRG2a}
K_{\sigma}&>&-K_{\rho}/2+\sqrt{K_{\rho}^2/4+4} .
\end{eqnarray}
As mentioned already in the main text, to make the two-particle scattering dominant, it is sufficient to either satisfy Eq.~(\ref{Eq:supRG1a}) or (\ref{Eq:supRG2a}). While Eq.~(\ref{Eq:supRG2a}) can only be satisfied for $K_{0}<1/\sqrt{3} \approx 0.58$, Eq.~(\ref{Eq:supRG1a}) can in principle tolerate slightly larger values of $K_0$. This can be understood by a careful analysis of intra- and inter-edge interaction terms. Evidently, $K_0<K_{\rho}$ since we obtain two additional (inter-edge) interaction channels in the constriction. They can be written as
\begin{eqnarray}
g_{2\parallel}(\rho_{R,\uparrow}(x)\rho_{L,\uparrow}(x)+\rho_{R,\downarrow}(x)\rho_{L,\downarrow}(x)),\\
g_{4\perp}(\rho_{R,\uparrow}(x)\rho_{R,\downarrow}(x)+\rho_{L,\uparrow}(x)\rho_{L,\downarrow}(x))
\end{eqnarray}
parameterized by coupling constants $g_{2\parallel}$ and $g_{4\perp}$. For fixed intra-edge interaction parameter $K_0$, the interaction parameters in the constriction can be written as
\begin{eqnarray}
\label{Eq:Krho}
K_{\rho}&=&\sqrt{\frac{\tilde{g}_{4\perp}-\tilde{g}_{2\parallel}+(2+\tilde{g}_{4\perp}-\tilde{g}_{2\parallel})K_0^2}{2+\tilde{g}_{2\parallel}+\tilde{g}_{4\perp}+(\tilde{g}_{2\parallel}+\tilde{g}_{4\perp})K_0^2}},\\
\label{Eq:Ksigma}
K_{\sigma}&=& \sqrt{\frac{2-\tilde{g}_{2\parallel}-\tilde{g}_{4\perp}-(\tilde{g}_{2\parallel}+\tilde{g}_{4\perp})K_0^2}{\tilde{g}_{2\parallel}-\tilde{g}_{4\perp}+(2+\tilde{g}_{2\parallel}-\tilde{g}_{4\perp})K_0^2}}
\end{eqnarray}
with $\tilde{g}_{\nu}=g_{\nu}/(2\pi v_F)$. Inserting Eqs.~(\ref{Eq:Krho}) and (\ref{Eq:Ksigma}) into Eq.~(\ref{Eq:supRG1a}), this yields a relation for $g_{2\parallel}$ and $g_{4\perp}$ that depends on the initial value of $K_0$. We obtain that there is a parameter space, for which Eq.~(\ref{Eq:RG1a}) is fulfilled, up to $K_{0}\sim 0.65$ provided that inter-edge interaction is smaller than intra-edge interaction, i.e. $g_{2\parallel},~g_{4\perp}\leq g_2$.

\subsection{Derivation of the hard-wall boundary conditions}

Here, we discuss the boundary conditions induced by an impurity in the constriction. With the corresponding Hamiltonian given by Eq.~(\ref{Eq:Imp}), we start with the non-interacting single-particle (first quantized) Schr\"odinger equation
\begin{equation}
\label{Eq:supSchrdinger}
\hat{h}\Psi(x)=\omega\Psi(x),
\end{equation}
where $\hat{h}=-iv_F\partial_x\tau_z\sigma_z+V\delta(x) \tau_x\sigma_0$ with $\tau_i$ and $\sigma_i$ ($i\in\{0,x,y,z\}$) Pauli matrices acting in edge- and spin-space, respectively. The basis is given by $\Psi(x)=[\psi_{R,\uparrow}(x),\psi_{L,\downarrow}(x),\psi_{L,\uparrow}(x),\psi_{R,\downarrow}(x)]^T$. Eq.~(\ref{Eq:supSchrdinger}) can be rearranged into
\begin{equation}
\partial_x\Psi(x)=\frac{i}{v_F}\tau_z\sigma_z\left[\omega-V\delta(x)\tau_x\sigma_0\right]\Psi(x),
\end{equation}
which is solved for the wave functions $\Psi(x)$ by integration \cite{Timm2012}
\begin{equation}
\label{Eq:supWave}
\Psi(x)=\hat{S}_{\leftarrow}\exp\bigg\lbrace\int_{x_0}^x\mathrm{d}x'\frac{i}{v_F}\tau_z\sigma_z\left[\omega-V\delta(x)\tau_x\sigma_0\right]\bigg\rbrace\Psi(x_0)
\end{equation}
with the spatial ordering operator $\hat{S}_{\leftarrow}$ that orders operators according to their spatial coordinate increasing from right to left. Eq.~(\ref{Eq:supWave}) implies that
\begin{eqnarray}
\label{Eq:supWave1}
\Psi(0^-)=\left[\cosh[V/v_F]+\tau_y\sigma_z\sinh[V/v_F]\right]\Psi(0^+).
\end{eqnarray}
Division of Eq.~(\ref{Eq:supWave1}) by $\cosh[V/v_F]$ and taking the limit $V\rightarrow\infty$, this yields the boundary condition for the wavefunction
\begin{equation}
[1+\tau_y\sigma_z]\Psi(0^+)=0
\end{equation}
that translates into two linearly independent conditions
\begin{eqnarray}
\label{Eq:supBC2}
\psi_{L,\!\uparrow}(0)\!=\!-i\psi_{R,\!\uparrow}(0),~~\psi_{R,\downarrow}(0)\!=i\psi_{L,\downarrow}\!(0).
\end{eqnarray}
Continuity of the wavefunction together with Eq.~(\ref{Eq:supBC2}), this fixes the form of the wavefunctions in region I (and right of that) in Fig.~\ref{Fig:system}(b) of the main text. In particular, as a result of the backscattering off the impurity, right- and left-moving particles in the two edges are no longer independent. Indeed, instead of having four independent wave functions $\psi_{r,\sigma}(x)$, the boundary conditions restrict us to the two valid solutions $\chi_{R,q}(x)=(e^{iqx},0,-ie^{-iqx},0)^T$ and $\chi_{L,q}(x)=(0,e^{-iqx},0,ie^{iqx})^T$. Consequently, the electronic (spinor) field operator $\hat{\Psi}(x)$ can, thus, be expanded in the basis $\lbrace \chi_{\nu,q}\rbrace$ as $\hat{\Psi}(x)=\sum_{q}[\chi_{R,q}(x)\hat{c}_{R,q}+\chi_{L,q}\hat{c}_{L.q}]$,with the fermionic annihilation operators $\hat{c}_{R,q},~\hat{c}_{L,q}$. To each of them we can assign a fermionic field $\hat{\psi}_{\nu}(x)=\sum_q e^{i\nu qx}\hat{c}_{\nu,q}$ with $\nu\in R,L$. Hence, the spinor Fermi field reads \cite{Fabrizio1995}
\begin{equation}
\label{Eq:supBC3}
\hat{\Psi}(x)=[\hat{\psi}_{R}(x),\hat{\psi}_L(x),-i\hat{\psi}_R(-x),i\hat{\psi}_L(-x)]^T.
\end{equation}
Adapting bosonization, Eq.~(\ref{Eq:supBC3}) implies that the number of Klein factors is reduced by half due to the boundary conditions. Thus, on the basis of the bosonization identity, Eq.~(\ref{Eq:FermiField}) of the main text, together with Eq.~(\ref{Eq:supBC3}), the corresponding relation for the bosonic fields of upper and lower edge are given by
\begin{eqnarray}
\label{Eq:supBC4a}
\phi_2(x)&=&-\phi_1(-x)-\frac{\pi}{2},\\
\label{Eq:supBC4b}
\theta_2(x)&=&\tilde{\theta}_1(-x).
\end{eqnarray}

\end{document}